\documentclass[12pt,showpacs,aps]{revtex4}
\usepackage{amsmath}
\usepackage{amssymb}
\usepackage{graphicx}
\usepackage{epsfig}
\usepackage{color}
\usepackage{ulem}
\usepackage{cancel}
\def\comment#1{}
\def\beq{\begin{equation}}
\def\eeq{\end{equation}}
\def\bea{\begin{eqnarray}}
\def\eea{\end{eqnarray}}
\def\no{\nonumber}
\def\red#1{\textcolor{red}{#1}}

\begin{document}

\title{High-energy neutrino emission in gravitational collapses}

\author{E.~Bavarsad,$^{1,2}$ 
and S.~S.~Xue$^{1}$}

\affiliation{$^{1}$ICRANet, P.zza della Repubblica 10, I--65122
Pescara, Physics Department and ICRA, University
of Rome, {\it La Sapienza} 
Rome, Italy\\
$^{2}$Department of Physics, Isfahan University of Technology,
Isfahan 84156-83111, Iran}

\date{Revised version \today}
\begin{abstract}
In this article, we present a study of high-energy neutrino emission in gravitational collapse. A compact star is treated as a complete degenerate Fermi gas of neutrons, protons and electrons. In gravitational collapse, its density reaches the thresholds for muon and pion productions, leading to high-energy neutrinos production. By using adiabatic approximation that macroscopic collapsing processes are much slower than microscopic processes of particle interactions, we adopt equilibrium equations of microscopic processes to obtain the number of neutrino productions. Assuming 10\% of variation in gravitational binding energy converted to the energy of produced neutrinos, we obtain fluxes of 10\,MeV electron-neutrinos and GeV electron and muon neutrinos. In addition, we compute the ratio ($< 1$) of total muon neutrino number to the total electron neutrino number at the source and at the Earth considering neutrino oscillations. We approximately obtain the number of GeV antineutrino events ($\gtrsim 1$) in an ordinary detector such as Kamiokande and total energy of neutrino flux ($\gtrsim 10^{53}\,\rm erg$), as a function of collapsing star mass.
\end{abstract}


\maketitle

\section{\label{sec:int}introduction}

When the most of nuclear fuel of a star has been consumed, there is nothing that resists against the inward gravitational pressure and collapse occurs. At the end of gravitational collapse will born a compact star which can be white dwarf, neutron star or black hole. In white dwarf degenerate Fermi gas of electrons supplies the equilibrium pressure and in the neutron stars they are supported by pressure of degenerate Fermi gas of neutrons. The collapse and supernova phase is thought to be accompanied by the most energetic neutrino burst as a neutron star or black hole is formed \cite{Burrows:1987zz}, see reviews \cite{Arnett:1990au, Bethe:1990mw, Burrows:1990ts}. This picture has been confirmed since the neutrino burst with mean energy of about 10\,MeV \cite{Burrows:1987zz}, known as SN1987A, was detected by both the Kamiokande-II \cite{Hirata:1987hu} and the IMB \cite{Bionta:1987qt} water Cerenkov detectors. In SN1987A a neutron star, whose mass is about $1.4\,M_{\odot}$, formed \cite{Arnett:1976dh, Cooperstein:1988zz}. However, stars with the mass $M\geq3.2\,M_{\odot}$ gravitationally collapse to black holes \cite{Rhoades:1974fn}. Recently \cite{Sekiguchi:2011zd}, numerical simulations for the merger of binary neutron stars are performed incorporating a finite-temperature, Shen's, equation of state and neutrino cooling effect. They showed that for such system a hypermassive neutron star results and black hole is not promptly formed after the onset of the merger as long as the total mass of the system is smaller than $3.2\,M_{\odot}$. The neutrino luminosity of the hypermassive neutron star was shown to be  $\sim(3-10)\times10^{53}\,\rm erg/s$. As opposed to the thermal 10\,MeV neutrinos, astrophysical objects like supernova core collapses \cite{Razzaque:2004yv, Razzaque:2005bh} supernova remnants \cite{Costantini:2004ap}, magnetars \cite{Zhang:2002xv}, neutron stars and pulsars \cite{Eichler:1978zp, Helfand:1979iv} can produce high-energy ($\geq1\,\rm GeV$) neutrinos, see reviews \cite{Dermer:2006xt, Stanev:2005kk, Becker:2007sv}. The production mechanism of high-energy neutrinos is the decay chain of charged pions and muons
\begin{eqnarray}\label{pd}
\pi^{-}&\rightarrow&\mu+\bar{\nu}_{\mu}\rightarrow\rm e+\bar{\nu}_{\mu}+\nu_{\mu}+\bar{\nu}_{\rm e},\no\\
\pi^{+}&\rightarrow&\mu^{+}+\nu_{\mu}\rightarrow\rm e^{+}+\bar{\nu}_{\mu}+\nu_{\mu}+\nu_{\rm e},
\end{eqnarray}
these pions are produced by accelerated protons \cite{Blasi:2000xm, Arons:2002yj} to high energies in the system via the photon-hadron interactions \cite{Mucke:1999yb} and inelastic hadronic processes \cite{Machner:1999ky}
\begin{eqnarray}
\rm p+\gamma&\rightarrow&\Delta^{+}\rightarrow\rm n+\pi^{+},\label{pgama}\\
\rm p+p&\rightarrow&\rm p+n+\pi^{+},\label{pp}
\end{eqnarray}
where $\gamma$ is the target photon. The most dominant channel for pion production in Eq.~(\ref{pgama}) is near the $\Delta$ resonance \cite{Mucke:1999yb}. Kilometer-scale neutrino detectors such as IceCube are constructed to observe these high-energy neutrinos \cite{Halzen:2006mq}.
\par
In this work we study another source of high-energy neutrinos from gravitationally collapsing stars to black holes. We compute number and mean energy of produced neutrinos in collapsing stars to black holes. To compute these quantities one has to solve rate equations which incorporate both, macroscopic hydrodynamic gravitational collapse process effects and microscopic particle interactions which produce neutrinos. However, time scale of particle interactions is much shorter than time scale of collapsing processes. As a result, gravitational collapse processes can be considered adiabatic in comparison with particle interaction processes. Based on these argumentations, we divide collapse process into infinitesimal steps where in each step the system is in its thermodynamic equilibrium state, which we call the adiabatic approximation \cite{mrx2012}. In addition to these thermodynamic equilibrium equations there are conserved quantities, which are total electric charge number, total baryon number and total lepton number of each flavor \cite{Shapiro}. We consider the collapsing star as a complete degenerate Fermi gas of neutrons, protons and electrons because the temperature is much smaller than their Fermi energies \cite{Shapiro, Weinberg}. Solving equilibrium equations analytically, we obtain the number of each spices of particles as a function of collapsing star density. When collapse proceeds density of star reaches to the amount that muon then pion can be produced in the system. As a result, in addition to the electron neutrinos which are already produced by $\beta$-processes, in companion to the muon production, muon neutrinos are produced.
Using number of produced neutrinos and assumption of 10\% \cite{Pagliaroli:2008ur} of gravitational binding energy converted to the energy of produced neutrinos, we compute the mean energy of neutrinos. We study neutrino oscillation effects on the primary flux, and compute the number of antineutrino events in an ordinary detector such as Kamiokande. The total energy of neutrino flux is another relevant quantity to observations and we compute it in this article.
\par
The paper is organized as follows: In Sec.~\ref{sec:stat}, we solve equilibrium equations and obtain number of each species of particles as a function of star density. In Sec.~\ref{sec:nu}, number of produced neutrinos is computed. In Sec.~\ref{sec:enrgy}, we compute mean energy of produced neutrinos. In Sec.~\ref{sec:osc}, we compute the ratio of total muon neutrinos to the total electron neutrinos at the source and at the Earth, considering neutrino oscillation effects; and the number of antineutrino events in an ordinary detector such as Kamiokande obtained, as well as total energy of neutrino flux. Finally in the last section we summery and discuss the results.

\section{\label{sec:stat}equilibrium states}

In this paper, we approximately consider compact stars, white dwarf and neutron stars, 
as a complete degenerate Fermi gas of neutrons, protons and electrons for the reason that the temperature is much smaller than their Fermi energies. In equilibrium state, each reaction between particles is balanced by its inverse reaction and leads to the chemical equilibrium condition \cite{Shapiro, Weinberg}
\begin{eqnarray}\label{cheqil}
\sum_{i=\rm particles}\mu_{i}dY_{i}=0,
\end{eqnarray}
where $\mu_i$ is the chemical potential of the particle-$i$, and $Y_{i}=n_{i}/n_{\rm B}$ is the concentration of the $i$-species of particle with respect to the baryon number density $n_{\rm B}$ such that $n_{i}$ is the number density of the particle-$i$. Therefore in equilibrium state the concentrations of the particles are not independent of each other, they are governed by the chemical equilibrium condition (\ref{cheqil}). In addition, total electric charge number, total baryon number $A$ and total lepton number of each flavor, must be conserved. Suppose that high-energy neutrinos are allowed to escape from the system, 
neutrino number densities and consequently their Fermi energies can be approximately set to zero in Eq.~(\ref{cheqil}) for equilibrium state.
Equation (\ref{cheqil}), total charge and baryon number conservations, are sufficiently determine all numbers of particles in equilibrium states.

\subsection{\label{sec:bta}$\beta$-processes}

\begin{figure}
\includegraphics[width=4.9in]{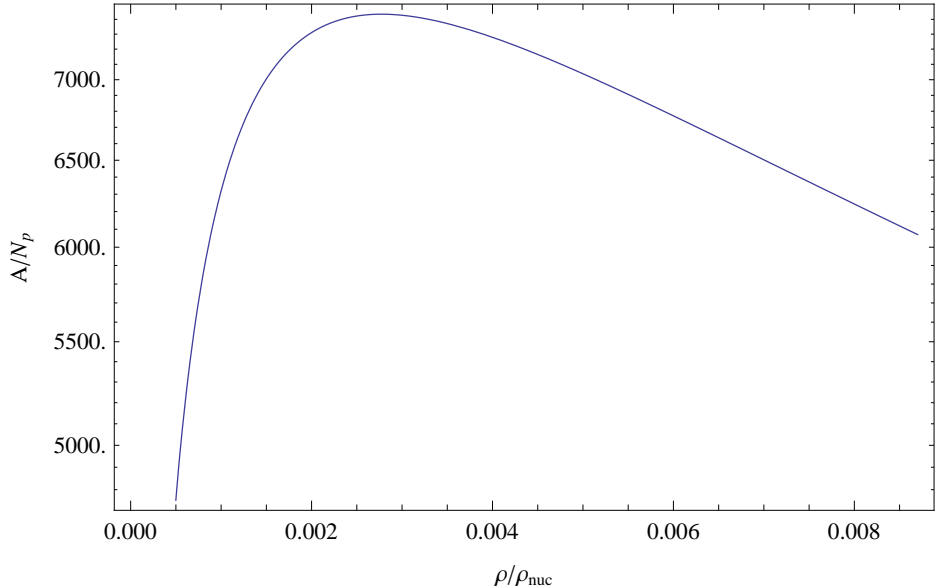}\\
\caption{The quantity $A/N_{\rm p}$ is plotted as a function of the stellar density $\rho$ in unit of nuclear density $\rho_{\rm nuc}$.}
\label{fig:ap}
\end{figure}

In the density region above the white dwarf density $\rho_{\rm wd}=8.73\times10^{-8}\rho_{\rm nuc}$, where the nuclear density is $\rho_{\rm nuc}=2.8\times10^{14}\,\rm gr\,cm^{-3}$, electrons are ultra relativistic and have enough energy to overcome the threshold
\begin{eqnarray}\label{btathr}
E_{\rm F,e}\geq(m_{\rm n}-m_{\rm p}),
\end{eqnarray}
then inverse $\beta$-decay reaction
\begin{eqnarray}\label{invbta}
\rm e+p\rightarrow\rm n+\nu_{e},
\end{eqnarray}
can proceed. In equilibrium state, the inverse $\beta$-decay reaction is balanced by its back reaction i.e., $\beta$-decay reaction
\begin{eqnarray}\label{btadcy}
\rm n\rightarrow\rm e+p+\bar{\nu}_{e}.
\end{eqnarray}
Variations of particle numbers due to the reactions (\ref{invbta}) and (\ref{btadcy}) respectively are given by
\begin{eqnarray}
dN_{\nu_{\rm e}}&=&dN_{\rm n}=-dN_{\rm p}=-dN_{\rm e},\label{invnum}\\
dN_{\bar{\nu}_{\rm e}}&=&dN_{\rm p}=dN_{\rm e}=-dN_{\rm n},\label{btvnum}
\end{eqnarray}
where $N_{i};\,\,\,\,\,i=\rm n,p,e,\nu_{\rm e}$ and $\bar{\nu}_{\rm e}$, respectively are total numbers of neutrons, protons, electrons, electron-neutrinos and
electron-antineutrinos. Thus the chemical equilibrium condition (\ref{cheqil}) requires following relation between chemical potentials of particles
\begin{eqnarray}\label{btaeq}
\mu_{\rm n}=\mu_{\rm p}+\mu_{\rm e}.
\end{eqnarray}
For a complete degenerate system at zero temperature $T$, chemical potentials $\mu_{i}=\sqrt{m_{i}^{2}+p_{{\rm F},i}^{2}}$, and Fermi momentum $p_{{\rm F},i}$  are known as functions of particle number densities $n_{i}$,
\begin{eqnarray}\label{fermom}
p_{{\rm F},i}(\rho)&=&(3\pi^{2}n_{i})^{1/3}\no\\
&=&\xi^{-1/2}\left(\frac{N_{i}}{A}\right)^{1/3}
\left(\frac{\rho}{\rho_{\rm nuc}}\right)^{1/3}m_{\pi},
\end{eqnarray}
where we adopt the approximation for spatial homogeneity and the numerical coefficient $\xi=0.173$. Total charge neutrality and conservation of total baryon
number respectively are given by
\begin{eqnarray}
N_{\rm p}&=&N_{\rm e},\label{qbta}\\
A&=&N_{\rm n}+N_{\rm p}.\label{bryn}
\end{eqnarray}
We obtain the solutions $N_{\rm p}$, $N_{\rm e}$ and $N_{\rm n}$ to Eqs.~(\ref{btaeq}),
(\ref{qbta}) and (\ref{bryn})
\begin{eqnarray}
N_{\rm p}(\rho)&\simeq&\frac{1}{8}A\left[1+
\xi\left(\frac{m_{\rm n}}{m_{\pi}}\right)^{2}
\left(\frac{\rho}{\rho_{\rm nuc}}\right)^{-2/3}\right]^{-3/2}\no\\
&\times&\left[\left(1+\xi\left(\frac{m_{\rm n}^{2}-m_{\rm p}^{2}
-m_{\rm e}^{2}}{m_{\pi}^{2}}\right)
\left(\frac{\rho}{\rho_{\rm nuc}}\right)^{-2/3}\right)^{2}
-4\xi^{2}\left(\frac{m_{\rm p}m_{\rm e}}{m_{\pi}^{2}}\right)^{2}
\left(\frac{\rho}{\rho_{\rm nuc}}\right)^{-4/3}
\right]^{3/2},\label{np}\no\\\\
N_{\rm e}(\rho)&=&N_{\rm p}(\rho),\label{ne}\\
N_{\rm n}(\rho)&=&A-N_{\rm p}(\rho),\label{nn}
\end{eqnarray}
such that in obtaining Eq.~(\ref{np}), only leading order terms i.e., $(N_{\rm p}/A)^{2/3}$ are taking into account and higher order terms are neglected.
Similar formulas to Solutions (\ref{np})-(\ref{nn}) can be found in \cite{Shapiro, Weinberg}. Solutions (\ref{np})-(\ref{nn}) completely determine the numbers of protons, electrons and neutrons as a function of the density $\rho$ of compact stars. In Fig.~\ref{fig:ap}, we plot the ratio $A/N_{\rm p}$ as a function of the density $\rho$, which is an important quantity to show charged compositions of compact stars at the density $\rho$. As shown in this figure, there are two main phases for neutrino productions by $\beta$-processes. In the first phase, $A/N_{\rm p}$ increases, through the inverse $\beta$-decay reaction (\ref{invbta}) electrons and protons having been converted into neutrons and  electron-neutrinos, the later escape from compact star. The ratio $A/N_{\rm p}$ reaches a maximum when the density reaches to the value
\begin{eqnarray}\label{dt}
\rho_{\rm T}\simeq 21.75\,\rho_{\rm nuc}\left[\frac{4((m_{\rm n}-m_{\rm p})^{2}
-m_{\rm e}^{2})}{m_{\rm n}^{2}}\right]^{3/4}\simeq2.8\times10^{-3}\rho_{\rm nuc},
\end{eqnarray}
which is called the transition density \cite{Shapiro, Weinberg}. In the second phase beyond the transition density (\ref{dt}) the ratio $A/N_{\rm p}$ decreases monotonically, through $\beta$-decay reaction (\ref{btadcy}) neutrons having been converted into protons, electrons and electron-antineutrinos escaping from compact star. We assume this second phase to end at the density at which other processes become important.

\subsection{\label{sec:mu}Production of muons}

\begin{figure}
\includegraphics[width=4.9in]{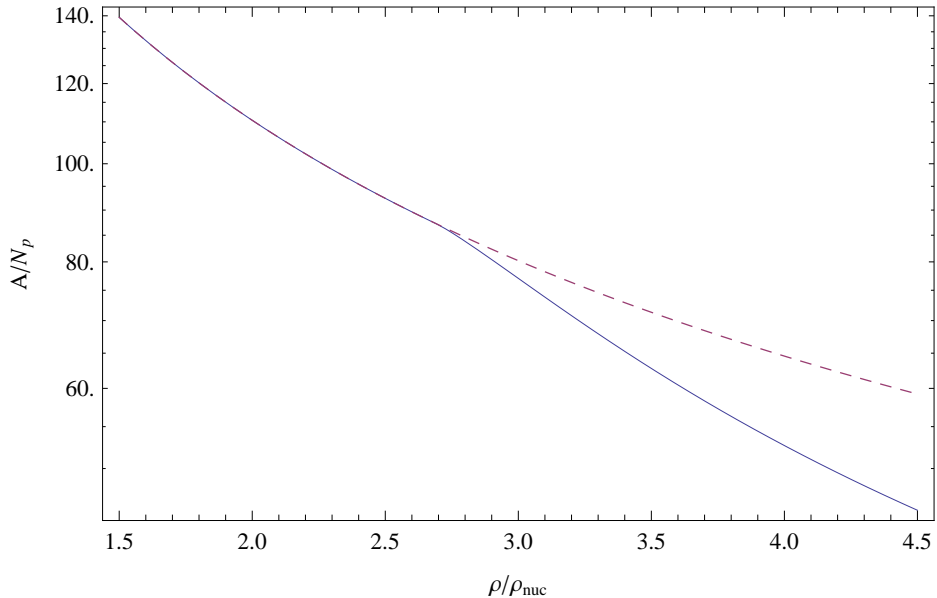}\\
\caption{The quantity $A/N_{\rm p}$ is plotted as a function of the density in the region $\rho\geq\rho_{\mu}=2.7\,\rho_{\rm nuc}$ in solid line.
Dashed line indicates the result obtained from $\beta$-processes.}
\label{fig:apy}
\end{figure}

When the stellar density $\rho$ reaches to the value $\rho_{\mu}\equiv 2.7\,\rho_{\rm nuc}$, the Fermi energy of electrons exceeds the muon mass, i.e., $E_{\rm F,e}\geq m_{\mu}$, then it is energetically favorable for electrons to turns into muons via
\begin{eqnarray}\label{muon}
\rm e+\bar{\nu}_{\rm e}\rightarrow\mu+\bar{\nu}_{\mu}.
\end{eqnarray}
And muons can decay to electrons
\begin{eqnarray}\label{mudcy}
\mu\rightarrow\rm e+\nu_{\mu}+\bar{\nu}_{\rm e},
\end{eqnarray}
so that muons and electrons are in equilibrium \cite{Shapiro}. The variations of particle numbers due to the reactions (\ref{muon}) and (\ref{mudcy}) respectively are given by
\begin{eqnarray}
dN_{\bar{\nu}_{\mu}}&=&dN_{\mu}=-dN_{\bar{\nu}_{\rm e}}=-dN_{\rm e},\label{muvnum}\\
dN_{\bar{\nu}_{\rm e}}&=&dN_{\nu_{\mu}}=dN_{\rm e}=-dN_{\mu},\label{mdvnum}
\end{eqnarray}
where $N_{i};\,\,\,\,\,i=\mu,\,\nu_{\mu}$ and $\bar{\nu}_{\mu}$, respectively are total number of muons, muon-neutrinos and muon-antineutrinos. The chemical equilibrium condition (\ref{cheqil}) gives the $\beta$-equilibrium condition (\ref{btaeq}) and the following relation between chemical potentials of electrons and muons \cite{Shapiro}
\begin{eqnarray}\label{mueq}
\mu_{\rm e}=\mu_{\mu}.
\end{eqnarray}
The conservations of total baryon numbers (\ref{bryn}) and total charge neutrality lead to
\begin{eqnarray}\label{qmu}
N_{\rm p}=N_{\rm e}+N_{\mu}.
\end{eqnarray}
In terms of the particle densities, Eqs.~(\ref{btaeq}), (\ref{bryn}), (\ref{mueq}) and (\ref{qmu}) form a sufficient set of equations to solve for finding particle numbers as a function of the stellar density. We obtain the solutions for the numbers of electrons, muons, protons and neutrons as functions of the stellar density $\rho$
\begin{eqnarray}
N_{\rm e}(\rho)&\simeq&\frac{1}{8}A\left[1+
\xi\left(\frac{m_{\rm n}}{m_{\pi}}\right)^{2}
\left(\frac{\rho}{\rho_{\rm nuc}}\right)^{-2/3}\right]^{-3/2}\no\\
&\times&\left[\left(1+\xi\left(\frac{m_{\rm n}^{2}-m_{\rm p}^{2}
-m_{\rm e}^{2}}{m_{\pi}^{2}}\right)
\left(\frac{\rho}{\rho_{\rm nuc}}\right)^{-2/3}\right)^{2}
-4\xi^{2}\left(\frac{m_{\rm p}m_{\rm e}}{m_{\pi}^{2}}\right)^{2}
\left(\frac{\rho}{\rho_{\rm nuc}}\right)^{-4/3}\right]^{3/2},\label{ney}\no\\\\
N_{\mu}(\rho)&=&A\left[\left(\frac{N_{\rm e}}{A}\right)^{2/3}-
\xi\left(\frac{m_{\mu}^{2}-m_{\rm e}^{2}}{m_{\pi}^{2}}\right)
\left(\frac{\rho}{\rho_{\rm nuc}}\right)^{-2/3}\right]^{3/2},\label{nm}\\
N_{\rm p}(\rho)&=&N_{\rm e}(\rho)+N_{\mu}(\rho),\label{npy}\\
N_{\rm n}(\rho)&=&A-N_{\rm p}(\rho),\label{nny}
\end{eqnarray}
such that in obtaining Eq.~(\ref{ney}), only leading order terms i.e., $(N_{\rm e}/A)^{2/3}$ are taking into account and the muon number $N_{\mu}$ are neglected in compared to the electron number $N_{\rm e}$. We plot the ratio $A/N_{\rm p}$ in Fig.~\ref{fig:apy}, and find it decreases faster than what is predicted by $\beta$-processes due to muon production.

\subsection{\label{sec:pi}Production of pions}

\begin{figure}
\includegraphics[width=4.9in]{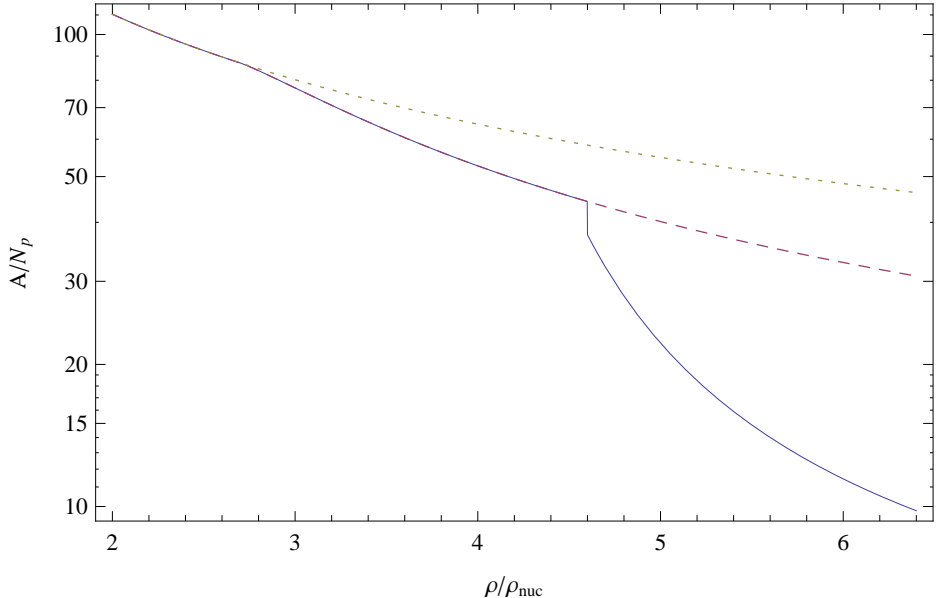}\\
\caption{The quantity $A/N_{\rm p}$ is plotted as a function of density in the region $\rho\geq\rho_{\pi}=4.6\,\rho_{\rm nuc}$ in solid line. Dotted and dashed lines respectively show what are predicted by $\beta$-processes and muon production.}
\label{fig:apz}
\end{figure}

If one ignores the effects of the strong interactions between pions and nucleons, the criterion for the formation of negatively charged pions in a compact star via the reaction
\begin{eqnarray}\label{pion}
\rm n\rightarrow p+\pi^{-},
\end{eqnarray}
is given by \cite{Shapiro}
\begin{eqnarray}\label{pieq}
\mu_{\rm n}-\mu_{\rm p}=\mu_{\pi}.
\end{eqnarray}
At low temperature $T\approx 0$, produced pions have approximately zero kinetic energy thus their chemical potential is close to the pion rest mass $\mu_{\pi}\approx m_{\pi}$, that determines the pion production threshold at the density $\rho_{\pi}\equiv4.6\,\rho_{\rm nuc}$. In equilibrium sate, produced pion can decay to muon which balanced by its back reaction
\begin{eqnarray}\label{pimu}
\pi^{-}\leftrightarrows\mu+\bar{\nu}_{\mu},
\end{eqnarray}
hence chemical equilibrium condition (\ref{cheqil}) requires
\begin{eqnarray}\label{pimueq}
\mu_{\pi}=\mu_{\mu}.
\end{eqnarray}
The chemical equilibrium condition leads to Eqs.~(\ref{pieq}) and (\ref{pimueq}), which are accompanied by the $\beta$-equilibrium (\ref{btaeq}) and electron and muon equilibrium (\ref{mueq}). As a usual in addition to the baryon number conservation (\ref{bryn}), charge neutrality leads to
\begin{eqnarray}\label{qpi}
N_{\rm p}=N_{\rm e}+N_{\mu}+N_{\pi},
\end{eqnarray}
where $N_{\pi}$ is the number of produced pions ($\pi^-$). By using expressions of chemical potentials in terms of particle densities, and solving Eqs.~(\ref{btaeq}), (\ref{bryn}), (\ref{mueq}), (\ref{pieq}), (\ref{pimueq}) and (\ref{qpi}), we obtain solutions for the numbers of electrons, muons, protons, neutrons and pions as a function of the stellar density $\rho$
\begin{eqnarray}
N_{\rm e}(\rho)&=&A\xi^{3/2}\left(\frac{\rho}{\rho_{\rm nuc}}\right)^{-1}
\left(1-\frac{m_{\rm e}^{2}}{m_{\pi}^{2}}\right)^{3/2},\label{nez}\\
N_{\mu}(\rho)&=&A\xi^{3/2}\left(\frac{\rho}{\rho_{\rm nuc}}\right)^{-1}
\left(1-\frac{m_{\mu}^{2}}{m_{\pi}^{2}}\right)^{3/2},\label{nmz}\\
N_{\rm p}(\rho)&\simeq&\frac{1}{\sqrt{8}}A\left[1+
\xi\left(1+\frac{m_{\rm n}^{2}-m_{\rm p}^{2}}{m_{\pi}^{2}}\right)
\left(\frac{\rho}{\rho_{\rm nuc}}\right)^{-2/3}\right]^{-3/2}\no\\
&\times&\left[\left(1-\xi\left(1-\frac{m_{\rm n}^{2}-m_{\rm p}^{2}
}{m_{\pi}^{2}}\right)
\left(\frac{\rho}{\rho_{\rm nuc}}\right)^{-2/3}\right)^{2}
-4\xi^{2}\left(\frac{m_{\rm p}}{m_{\pi}}\right)^{2}
\left(\frac{\rho}{\rho_{\rm nuc}}\right)^{-4/3}\right]^{3/2},\label{npz}\\
N_{\rm n}(\rho)&=&A-N_{\rm p}(\rho),\label{nnz}\\
N_{\pi}(\rho)&=&N_{\rm p}(\rho)-N_{\rm e}(\rho)-N_{\mu}(\rho),\label{npi}
\end{eqnarray}
such that in obtaining Eq.~(\ref{npz}), only leading order terms i.e., $(N_{\rm p}/A)^{2/3}$ are taking into account and higher order terms are neglected.
In Fig.~\ref{fig:apz}, we plot the radio $A/N_{\rm p}$ in this region $\rho\geq\rho_{\pi}$, and find the production of negatively charged pions causes a faster
decreasing of this ratio, comparing to the result obtained from $\beta$-processes and muon production.

\section{\label{sec:nu}neutrino production in gravitational collapses}

Stellar core density increases during gravitational collapses. As studied in the Sec.~\ref{sec:stat} particle numbers $N_{\rm n,p,e,\mu,\pi}$ are functions of the stellar core density $\rho$, thus change in collapses. These variations of particle numbers are due to the $\beta$-processes (\ref{invbta}), (\ref{btadcy}) and processes of muon-production (\ref{muon}), (\ref{mudcy}) and pion-production (\ref{pion}), (\ref{pimu}). Because the time-scales of these microscopic processes are much shorter than the time-scale of macroscopic processes, such as gravitational collapse and hydrodynamical processes, we approximately treat gravitational collapse and other macroscopic processes as adiabatic processes. Namely, gravitational collapse process can be divided into infinitesimal steps and at each step the system can be considered in its equilibrium state. Hence the results of the Sec.~\ref{sec:stat} are applicable at each step of gravitational collapse.
\par
\begin{figure}
\includegraphics[width=4.9in]{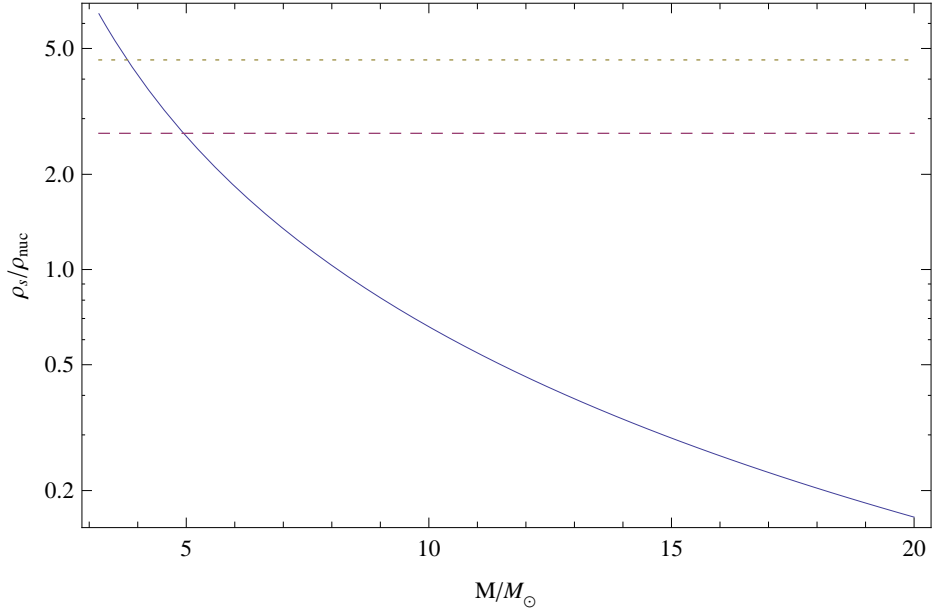}\\
\caption{The density $\rho$ at the Schwarzschild radius of the star $\rho_{\rm s}$, is plotted as a function of collapsing star mass $M$, in unit of Solar mass $M_{\odot}$, in solid line. Dashed and dotted lines respectively indicate the muon and pion production threshold densities $\rho_{\mu}$ and $\rho_{\pi}$.}
\label{fig:ds}
\end{figure}

Let $M$ be the mass of collapsing stars. When $M\geq3.2\,M_{\odot}$, black holes are formed at the Schwarzschild radius $R_{\rm s}=2GM$ of the star \cite{Ohanian}, where $G$ is gravitational constant. This indicates a maximal density $\rho_{\rm s}\equiv M/(\frac{4}{3}\pi R_{\rm s}^{3})$, which is plotted as a function of the mass $M$ of collapsing stellar cores in Fig.~\ref{fig:ds}, to compare with the density thresholds $\rho_\mu$ and $\rho_\pi$ of muon and pion productions discussed in the Sec.~\ref{sec:stat}. As a result, we conclude the following three cases
\begin{enumerate}
\item
$3.2\,M_{\odot}\leq M\leq3.8\,M_{\odot}$, the density thresholds $\rho_\mu$ and $\rho_\pi$ of both muon and pion productions can be reached;
\item
$3.9\,M_{\odot}\leq M\leq4.9\,M_{\odot}$, only density threshold $\rho_\mu$ of muon production can be reached;
\item
$M\geq 5\,M_{\odot}$ the density thresholds $\rho_\mu$ and $\rho_\pi$ of both muon and pion productions cannot be reached.
\end{enumerate}
According to these three cases and the density thresholds discussed in the Sec.~\ref{sec:stat}, we are ready to calculate the numbers of neutrino produced in different phases.

\subsection{\label{sec:a}Phase $a$}

In the phase $a$ from white dwarf density $\rho_{\rm wd}$ to the transition density $\rho_{\rm T}$ (\ref{dt}), Eq.~(\ref{invnum}) gives $dN_{\nu_{\rm e}}=-dN_{\rm e}$, thus the number of produced electron-neutrinos is
\begin{eqnarray}\label{nea}
N_{\nu_{\rm e}}=-\int_{\rho_{\rm wd}}^{\rho_{\rm T}}dN_{\rm e}=N_{\rm e}(\rho_{\rm wd})-N_{\rm e}(\rho_{\rm T})\simeq A/2,
\end{eqnarray}
where numbers of electrons, protons and neutrons are
\begin{eqnarray}\label{nwd}
N_{\rm e}(\rho_{\rm wd})=N_{\rm p}(\rho_{\rm wd})\simeq N_{\rm n}(\rho_{\rm wd})
\simeq A/2,
\end{eqnarray}
at the white dwarf density $\rho_{\rm wd}$, and Fig.~\ref{fig:ap} shows
\begin{eqnarray}\label{nt}
N_{\rm e}(\rho_{\rm T})=N_{\rm p}(\rho_{\rm T})\simeq 1.34\times10^{-4}A,
\end{eqnarray}
at the transition density $\rho_{\rm T}$.

\subsection{\label{sec:b}Phase $b$}

\begin{figure}
\includegraphics[width=4.9in]{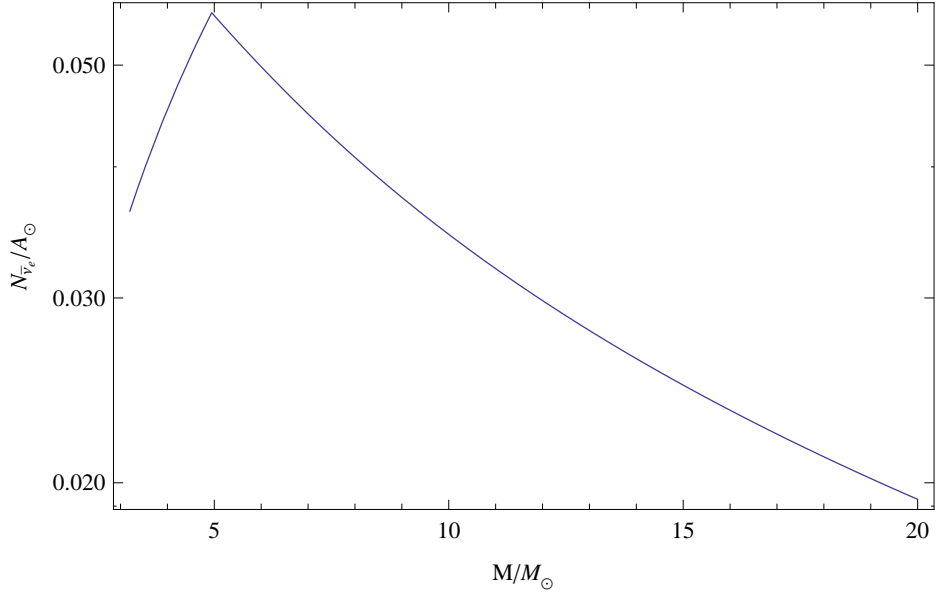}\\
\caption{The number of electron-antineutrinos $N_{\bar{\nu}_{\rm e}}$, in unit of Solar baryon number $A_{\odot}\simeq1.19\times10^{57}$, produced in the phase $b$, is plotted as a function of collapsing star mass $M$.}
\label{fig:nb}
\end{figure}
The phase $b$ of neutrino production begins from the transition density $\rho_{\rm T}$ to the density threshold $\rho_\mu$ for cases 1 and 2, and to the maximal density $\rho_{\rm s}$ for case 3. The number of produced electron-antineutrinos, due to the $\beta$-decay reaction (\ref{btadcy}), is equal to the variation of electron number $dN_{\bar{\nu}_{\rm e}}=dN_{\rm e}$ shown in Eq.~(\ref{btvnum}). In cases 1 and 2, we obtain the total number of produced electron-antineutrinos
\begin{eqnarray}\label{neb}
N_{\bar{\nu}_{\rm e}}=\int_{\rho_{\rm T}}^{\rho_{\mu}}dN_{\rm e}
=N_{\rm e}(\rho_{\mu})-N_{\rm e}(\rho_{\rm T})\simeq1.13\times10^{-2}A,
\end{eqnarray}
where $N_{\rm e}(\rho_{\mu})$ is given by Eq.~(\ref{ney}). In case 3, we obtain
\begin{eqnarray}\label{nebs}
N_{\bar{\nu}_{\rm e}}=\int_{\rho_{\rm T}}^{\rho_{\rm s}}dN_{\rm e}=N_{\rm e}(\rho_{\rm s})-N_{\rm e}(\rho_{\rm T}),
\end{eqnarray}
where $N_{\rm e}(\rho_{\rm s})$ is given by Eq.~(\ref{ne}). In Fig.~\ref{fig:nb}, we plot Eqs.~(\ref{neb}) and (\ref{nebs}), which shows $N_{\bar{\nu}_{\rm e}}$ increases for cases 1 and 2, and decreases as collapsing mass $M$ increases for case 3.

\subsection{\label{sub:t}Phase $c$}

\begin{figure}
\includegraphics[width=4.9in]{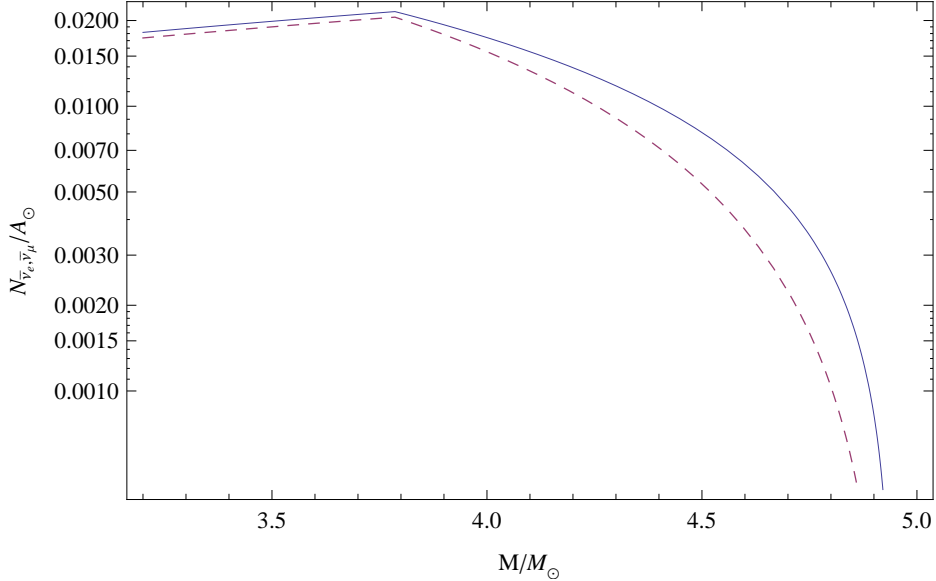}\\
\caption{The number of electron-antineutrinos $N_{\bar{\nu}_{\rm e}}$ and muon-antineutrinos $N_{\bar{\nu}_{\mu}}$ produced in the phase $c$ are plotted as a function of collapsing star mass $M$, respectively in solid and dashed lines.}
\label{fig:nc}
\end{figure}
The phase $c$ of neutrino production begins from muon production density $\rho_{\mu}$ to the density threshold $\rho_\pi$ for cases 1, and to the maximal density $\rho_{\rm s}$ for case 2. In this phase in addition to the $\beta$-processes, reactions (\ref{muon}) and (\ref{mudcy}) contribute to neutrino productions. Total number variations of electrons and muons, given by Eqs.~(\ref{ney}) and (\ref{nm}), are equal to the number of produced electron and muon antineutrinos, i.e.,
$dN_{\bar{\nu}_{\rm e}}=dN_{\rm e}$ and $dN_{\bar{\nu}_{\mu}}=dN_{\mu}$, due to the conservations of electron and muon flavor numbers. In case 1, we obtain the numbers of produced electron-antineutrinos $N_{\bar{\nu}_{\rm e}}$ and muon-antineutrinos $N_{\bar{\nu}_{\mu}}$
\begin{eqnarray}
N_{\bar{\nu}_{\rm e}}&=&\int_{\rho_{\mu}}^{\rho_{\pi}}dN_{\rm e}=N_{\rm e}(\rho_{\pi})-N_{\rm e}(\rho_{\mu})\simeq5.7\times10^{-3}A,
\label{nec}\\
N_{\bar{\nu}_{\mu}}&=&\int_{\rho_{\mu}}^{\rho_{\pi}}dN_{\mu}=N_{\mu}(\rho_{\pi})-N_{\mu}(\rho_{\mu})\simeq5.4\times10^{-3}A,
\label{nmc}
\end{eqnarray}
where we use Eqs.~(\ref{ney}) and (\ref{nm}). In case 2, we obtain the number of produced electron and muon antineutrinos
\begin{eqnarray}
N_{\bar{\nu}_{\rm e}}&=&\int_{\rho_{\mu}}^{\rho_{\rm s}}dN_{\rm e}=N_{\rm e}(\rho_{\rm s})-N_{\rm e}(\rho_{\mu}),
\label{necs}\\
N_{\bar{\nu}_{\mu}}&=&\int_{\rho_{\mu}}^{\rho_{\rm s}}dN_{\mu}=N_{\mu}(\rho_{\rm s})-N_{\mu}(\rho_{\mu}),
\label{nmcs}
\end{eqnarray}
where $N_{\rm e}(\rho_{\rm s})$ and $N_{\mu}(\rho_{\rm s})$ are given by Eqs.~(\ref{ney}) and (\ref{nm}). The results (\ref{nec})-(\ref{nmcs}) are plotted in Fig.~\ref{fig:nc}, which shows the numbers $N_{\bar{\nu}_{\rm e}}$ and $N_{\bar{\nu}_{\mu}}$ increase for case 1 and decrease as collapsing star mass $M$ increases for case 2.

\subsection{\label{sec:d}Phase $d$}

\begin{figure}
\includegraphics[width=4.9in]{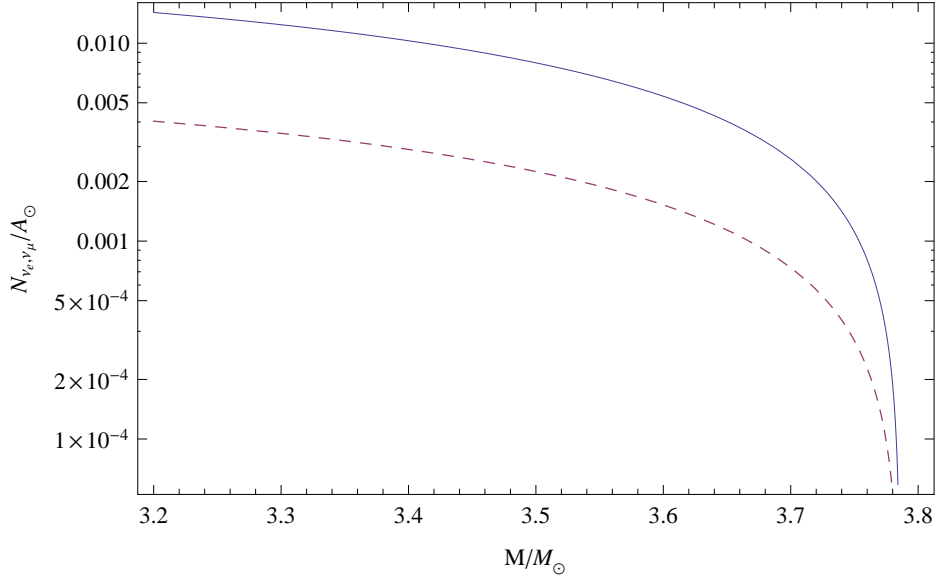}\\
\caption{The number of electron-neutrinos $N_{\nu_{\rm e}}$ and muon-neutrinos $N_{\nu_{\mu}}$ produced in the phase $d$ are plotted as a function of collapsing star mass $M$, respectively in solid and dashed lines.}
\label{fig:nd}
\end{figure}
The Phase $d$ of neutrino production begins from the threshold density $\rho_{\pi}$ for pion production and ends to the maximal density $\rho_{\rm s}$. This phase can only be reached in case 1. Equations (\ref{nez}) and (\ref{nmz}) show that by increasing the density $\rho$, electron number $N_{\rm e}$ and muon number $N_{\mu}$ decrease. Therefore electron and muon neutrinos are produced, $dN_{\nu_{\rm e}}=-dN_{\rm e}$ and $dN_{\nu_{\mu}}=-dN_{\mu}$, and the total numbers are
\begin{eqnarray}
N_{\nu_{\rm e}}&=&-\int_{\rho_{\pi}}^{\rho_{\rm s}}dN_{\rm e}=N_{\rm e}(\rho_{\pi})-N_{\rm e}(\rho_{\rm s}),
\label{ned}\\
N_{\nu_{\mu}}&=&-\int_{\rho_{\pi}}^{\rho_{\rm s}}dN_{\mu}=N_{\mu}(\rho_{\pi})-N_{\mu}(\rho_{\rm s}),
\label{nmd}
\end{eqnarray}
which are plotted by Fig.~\ref{fig:nd}, we find the number of neutrino productions are decreasing functions of collapsing star mass $M$.

\section{\label{sec:enrgy}mean energy of neutrinos}

\begin{figure}
\includegraphics[width=4.9in]{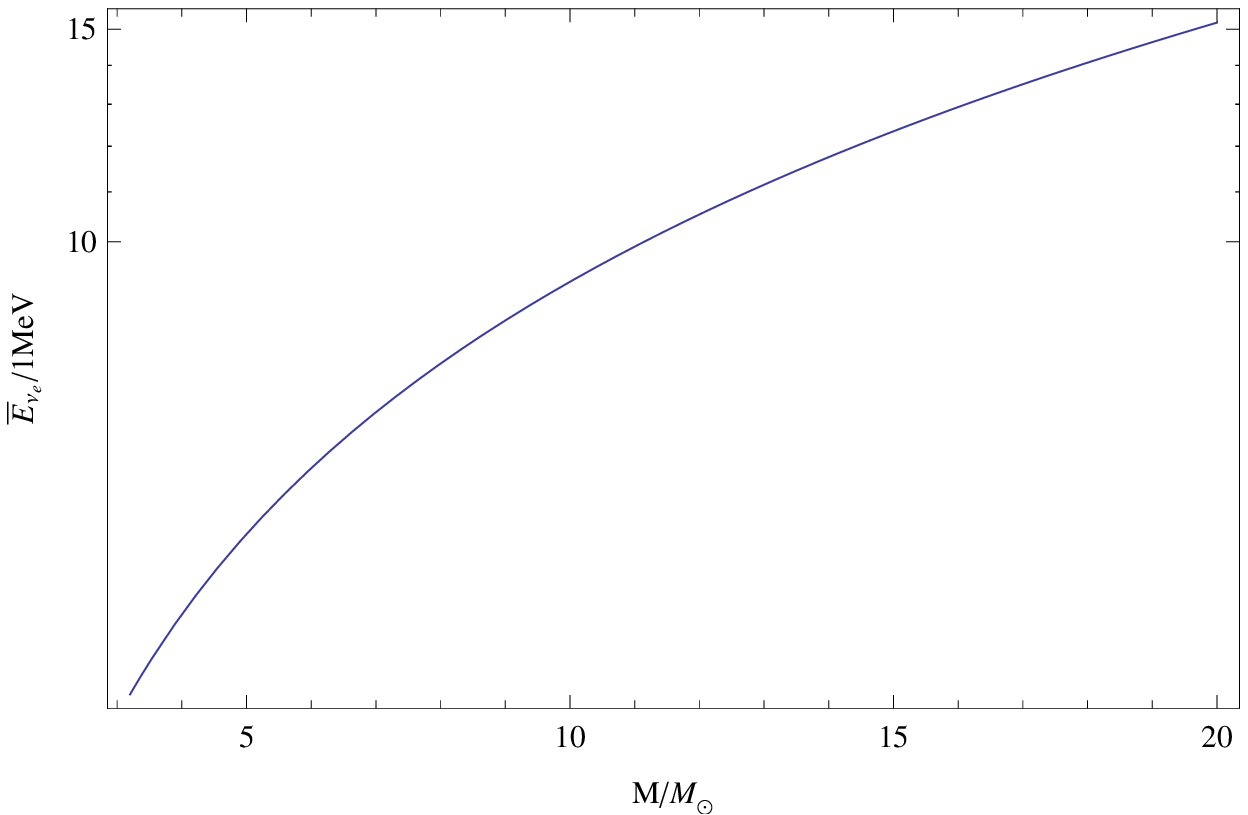}\\
\caption{Mean energy $\overline{E}_{\nu_{\rm e}}$, in unit of MeV, of produced electron-neutrinos in the phase $a$ is plotted as a function of collapsing star mass $M$.}
\label{fig:ea}
\end{figure}
The gravitational energy of stellar core with mass $M$ and radius $R$ is given by
\begin{eqnarray}\label{gravty}
U=\int_{0}^{R}4\pi r^{2}\left[1-\left(1-\frac{2G\mathcal{M}(r)}{r}\right)^{-1/2}\right]\rho(r)dr,
\end{eqnarray}
where $\mathcal{M}(r)\equiv\int_0^r 4\pi r'^2\rho(r')dr'$ \cite{Weinberg}. Assuming uniform density $\rho=M/(\frac{4}{3}\pi R^{3})$, we have
\begin{eqnarray}\label{grapot}
U=M\left[1-\frac{3}{2}a^{-3}\left(\arcsin(a)-a\sqrt{1-a^{2}}\right)\right];
\hspace{1cm}a\equiv\sqrt{\frac{2GM}{R}}.
\end{eqnarray}
During collapses from radius $R$ to $R-dR$, the variation of gravitational energy
\begin{eqnarray}\label{vargrv}
dU=\frac{3M}{4R}\frac{1}{\sqrt{1-a^{2}}}\left[1+3a^{-3}\left(\sqrt{1-a^{2}}\arcsin(a)-a\right)\right]dR,
\end{eqnarray}
is converted to the kinetic energy and internal energy of the particles.
\par
To compute the mean energy $\overline{E}_{\nu}$ of emitted neutrinos, we assume 10\% of total variation of gravitational energy $dU$ converted to the energy of produced neutrinos \cite{Pagliaroli:2008ur}. Suppose that $N_{\nu}$ neutrinos are emitted in gravitational collapse process from the radius $R_{\rm i}$ to the radius $R_{\rm f}$, observed mean energy of neutrinos can be calculated by
\begin{eqnarray}\label{difme}
\overline{E}_{\nu}=\frac{0.1}{N_{\nu}}\int_{R_{\rm i}}^{R_{\rm f}}Z(R)dU,
\end{eqnarray}
where $Z$ is the gravitational red shift factor
\begin{eqnarray}\label{red}
Z(R)=\sqrt{1-\frac{2GM}{R}}.
\end{eqnarray}
Using Eqs.~(\ref{vargrv})-(\ref{red}) and the number of produced neutrinos in each phases discussed in the Sec.~\ref{sec:nu}, we obtain mean energies of neutrinos.
\begin{figure}
\includegraphics[width=4.9in]{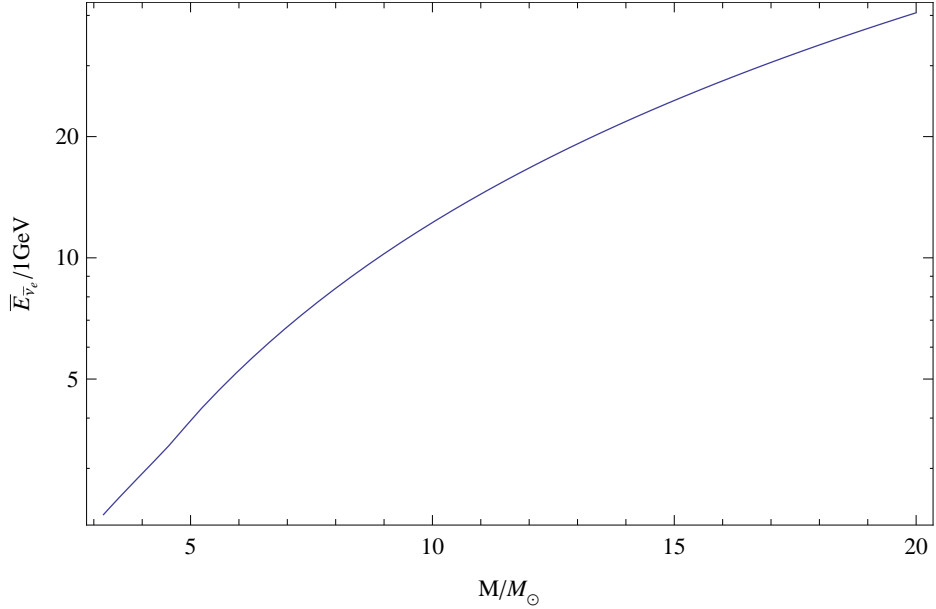}\\
\caption{Mean energy $\overline{E}_{\bar{\nu}_{\rm e}}$, in unit of GeV, of produced electron-antineutrinos in the phase $b$, is plotted as a function of collapsing star mass $M$.}
\label{fig:eb}
\end{figure}
In the phase $a$, the mean energy of produced $N_{\nu_{\rm e}}$, see Eq.~(\ref{nea}), electron-neutrinos is given by
\begin{eqnarray}\label{ea}
\overline{E}_{\nu_{\rm e}}=\frac{0.1}{N_{\nu_{\rm e}}}\int_{R_{\rm wd}}^{R_{\rm T}}Z(R)dU,
\end{eqnarray}
where $R_{\rm wd}$ is the radius at which the collapsing core density $\rho$ is equal to the white dwarf density $\rho_{\rm wd}$, and $R_{\rm T}$ is the radius at which the density reaches to the transition density $\rho_{\rm T}$, see Eq.~(\ref{dt}). We plot this result (\ref{ea}) in Fig.~\ref{fig:ea}, and find the neutrino mean energy is about 10\,MeV, which is in agreement with neutrino energy observed in supernova explosions.
In the phase $b$, the mean energy of produced electron-antineutrinos for cases 1 and 2 is given by
\begin{eqnarray}\label{eb}
\overline{E}_{\bar{\nu}_{\rm e}}=\frac{0.1}{N_{\bar{\nu}_{\rm e}}}\int_{R_{\rm T}}^{R_{\mu}}Z(R)dU,
\end{eqnarray}
where $N_{\bar{\nu}_{\rm e}}$ is calculated by Eq.~(\ref{neb}) and $R_{\mu}$ is a radius at which the density $\rho$ reaches to $\rho_{\mu}$. Whereas for case 3, Eq.~(\ref{eb}) becomes
\begin{eqnarray}\label{ebs}
\overline{E}_{\bar{\nu}_{\rm e}}=\frac{0.1}{N_{\bar{\nu}_{\rm e}}}\int_{R_{\rm T}}^{R_{\rm s}}Z(R)dU,
\end{eqnarray}
where $N_{\bar{\nu}_{\rm e}}$ is calculated by Eq.~(\ref{nebs}). The results (\ref{eb}) and (\ref{ebs}) are shown in Fig.~\ref{fig:eb}. Compared with results in the first phase, the electron-antineutrino mean energy about GeV is larger than the electron-neutrino mean energy, because in this phase the variation of gravitational energy is larger and the number of produced neutrino is smaller, see Fig.~\ref{fig:nb} and Eq.~(\ref{nea}).
\begin{figure}
\includegraphics[width=4.9in]{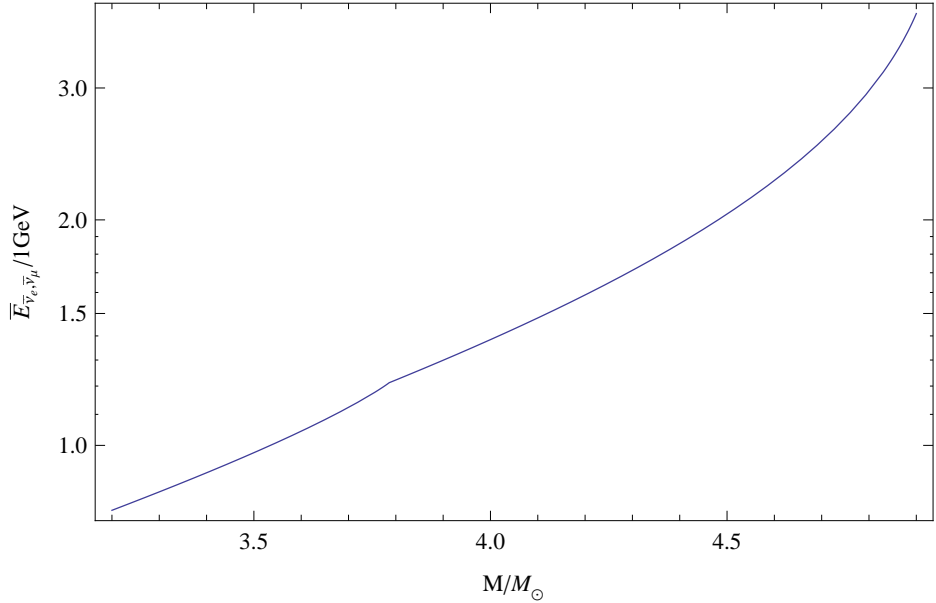}\\
\caption{Mean energy electron-antineutrinos and muon-antineutrinos $\overline{E}_{\bar{\nu}_{\rm e},\bar{\nu}_{\mu}}$, which are produced in the phase $c$, is plotted as a function of collapsing star mass $M$.}
\label{fig:ec}
\end{figure}
In the phase $c$, the mean energy of produced electron-antineutrinos and muon-antineutrinos for case 1 is given by
\begin{eqnarray}\label{ec}
\overline{E}_{\bar{\nu}_{\rm e},\bar{\nu}_{\mu}}=\frac{0.1}{(N_{\bar{\nu}_{\rm e}}+N_{\bar{\nu}_{\mu}})}\int_{R_{\mu}}^{R_{\pi}}Z(R)dU,
\end{eqnarray}
where $R_{\pi}$ is the radius at which the density $\rho$ reaches $\rho_{\pi}$, neutrino numbers $N_{\bar{\nu}_{\rm e}}$ and $N_{\bar{\nu}_{\mu}}$ are respectively given by Eqs.~(\ref{nec}) and (\ref{nmc}). Whereas for case 2 Eq.~(\ref{ec}) becomes
\begin{eqnarray}\label{ecs}
\overline{E}_{\bar{\nu}_{\rm e},\bar{\nu}_{\mu}}=\frac{0.1}{(N_{\bar{\nu}_{\rm e}}+N_{\bar{\nu}_{\mu}})}\int_{R_{\mu}}^{R_{\rm s}}Z(R)dU,
\end{eqnarray}
where neutrino numbers $N_{\bar{\nu}_{\rm e}}$ and $N_{\bar{\nu}_{\mu}}$ are respectively given by Eqs.~(\ref{necs}) and (\ref{nmcs}). The results (\ref{ec}) and (\ref{ecs}) are shown in Fig.~\ref{fig:ec}.
\begin{figure}
\includegraphics[width=4.9in]{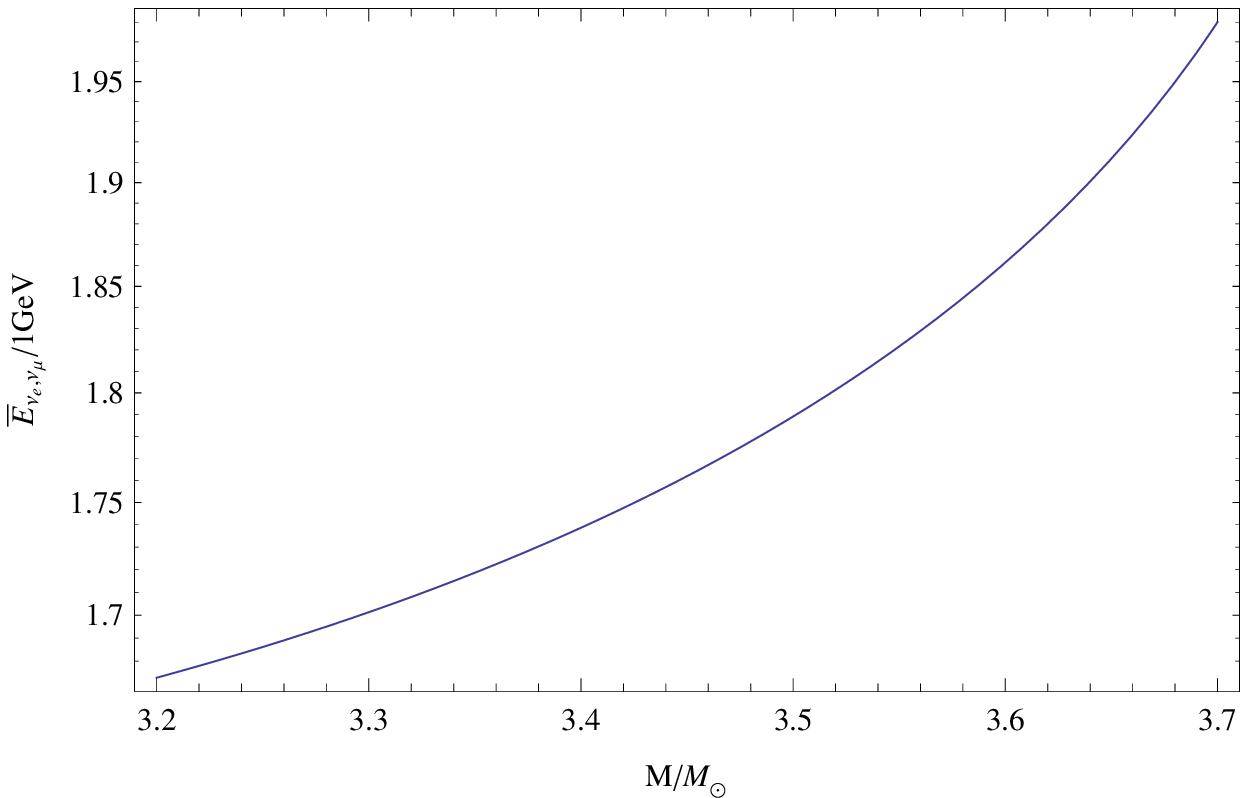}\\
\caption{Mean energy electron-neutrinos and muon-neutrinos $\overline{E}_{\nu_{\rm e},\nu_{\mu}}$, which are produced in the phase $d$, is plotted as a function of collapsing star mass $M$.}
\label{fig:ed}
\end{figure}
In the phase $d$, the mean energy of produced electron-neutrinos and muon-neutrinos is given by
\begin{eqnarray}\label{ed}
\overline{E}_{\nu_{\rm e},\nu_{\mu}}=\frac{0.1}{(N_{\nu_{\rm e}}+N_{\nu_{\mu}})}\int_{R_{\pi}}^{R_{\rm s}}Z(R)dU,
\end{eqnarray}
where the neutrino numbers $N_{\nu_{\rm e}}$ and $N_{\nu_{\mu}}$ are respectively given by Eqs.~(\ref{ned}) and (\ref{nmd}). The result (\ref{ed}) is shown in Fig.~\ref{fig:ed}. We find that the mean energy neutrinos produced via muon and pion productions is about GeV, see Figs.~\ref{fig:ec} and \ref{fig:ed}.
\begin{figure}
\includegraphics[width=4.9in]{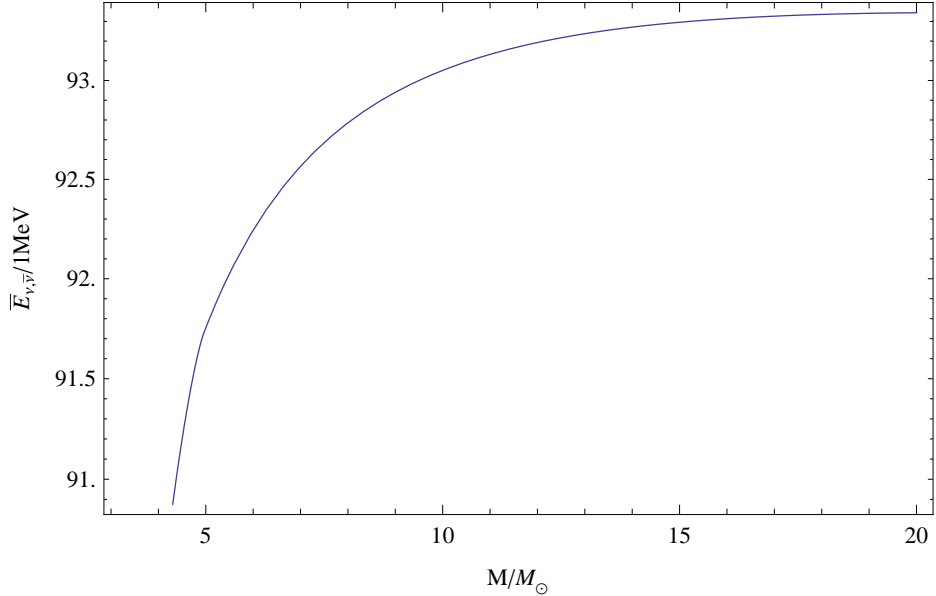}\\
\caption{Mean energy $\overline{E}_{\nu,\bar{\nu}}$ of total produced neutrinos and antineutrinos in the phases $a,b,c$ and $d$, is plotted as a function of collapsing star mass $M$.}
\label{fig:et}
\end{figure}
We also calculate the mean energy of total produced neutrinos throughout phases $a,b,c$ and $d$
\begin{eqnarray}\label{et}
\overline{E}_{\nu,\bar{\nu}}&=&\frac{0.1}{N_{\nu,\bar{\nu}}}\int_{R_{\rm wd}}^{R_{\rm s}}Z(R)dU;\no\\
N_{\nu,\bar{\nu}}&=&N_{\nu_{\rm e}}^{(a)}+N_{\bar\nu_{\rm e}}^{(b)}+N_{\bar\nu_{\rm e}}^{(c)}+N_{\bar\nu_{\mu}}^{(c)}+N_{\nu_{\rm e}}^{(d)}+N_{\nu_{\mu}}^{(d)},
\end{eqnarray}
where the superscripts show the related neutrino production phases. The result (\ref{et}) is shown in Fig.~\ref{fig:et}, and we find neutrino mean energy is about 90\,MeV, which is larger than neutrino mean energy 10\,MeV in supernova explosion. This is due to more gravitational energy converted to neutrino energy when stars gravitationally collapse to black holes.

\section{\label{sec:osc}oscillation and detection of neutrinos}

There are several observations, see \cite{Strumia:2006db} and references there in, indicate that neutrino has mass. Thus a neutrino with flavor $\ell$ is a superposition of the physical fields $\nu_{\alpha}$ with different masses $m_{\alpha}$
\begin{eqnarray}\label{mixing}
|\nu_{\ell}\rangle=\sum_{\alpha}U_{\ell\alpha}|\nu_{\alpha}\rangle,
\end{eqnarray}
where $U$ is unitary matrix which signifies neutrino mixing \cite{Mohapatra}. By considering a simple-minded approach to the propagation of this state, one can assume that the three-momentum $\vec{p}$ of the different components in the neutrino beam are the same. However, since their masses are different, the energies of all these components cannot be equal, and given by $E_{\alpha}=\sqrt{\vec{p}^{2}+m_{\alpha}^{2}}$. In the situation that we are studying, neutrinos are extremely relativistic, so that we can approximate the energy-momentum relation as $E_{\alpha}\simeq|\vec{p}|+m_{\alpha}^{2}/(2|\vec{p}|)$. After time $t$, or distance $L$, the evolution of the initial beam of Eq.~(\ref{mixing}) gives
\begin{eqnarray}\label{evul}
|\nu_{\ell}(t)\rangle=\sum_{\alpha}U_{\ell\alpha}e^{-iE_{\alpha}t}|\nu_{\alpha}\rangle.
\end{eqnarray}
Thus the probability of finding a $\nu_{\ell'}$ in the original $\nu_{\ell}$ beam, at any distance $L$, is
\begin{eqnarray}\label{prob}
P_{\nu_{\ell}\nu_{\ell'}}&=&|\langle\nu_{\ell'}|\nu_{\ell}(L)\rangle|^{2}\no\\
&=&\delta_{\ell\ell'}-4\sum_{\alpha>\beta}U_{\ell\alpha}U_{\ell'\alpha}U_{\ell\beta}U_{\ell'\beta}\sin^{2}
\left(\frac{\Delta m^{2}_{\alpha\beta}}{4E}L\right),
\end{eqnarray}
where we use the notation $|\vec{p}|=E$ and
\begin{eqnarray}\label{massqr}
\Delta m^{2}_{\alpha\beta}\equiv m^{2}_{\alpha}- m^{2}_{\beta}.
\end{eqnarray}
In writing Eq.~(\ref{prob}) CP-conservation has been assumed when the mixing matrix $U$ is real \cite{Mohapatra}. For three generations, the most general mixing matrix ignoring CP violation is given by
\begin{eqnarray}\label{mixmat}
U=
\left(
  \begin{array}{ccc}
    c_{12}c_{13} & -s_{12}c_{13} & s_{13} \\
    s_{12}c_{23}+c_{12}s_{23}s_{13} & c_{12}c_{23}-s_{12}s_{23}s_{13} & -s_{23}c_{13} \\
    s_{12}s_{23}-c_{12}c_{23}s_{13} & c_{12}s_{23}+s_{12}c_{23}s_{13} & c_{23}c_{13} \\
  \end{array}
\right),
\end{eqnarray}
where $c_{ij}=\cos\theta_{ij}$ and $s_{ij}=\sin\theta_{ij}$. Analysis of the most recent data \cite{Nakamura:2010zzi} on the oscillation parameters indicate that
\begin{eqnarray}\label{data}
\sin^{2}(2\theta_{12})&=&0.87\pm 0.03,\hspace{1cm}\Delta m^{2}_{21}=(7.59\pm 0.20)\times 10^{-5}{\rm eV}^{2},\no\\
\sin^{2}(2\theta_{23})&>&0.92,\hspace{2.25cm}\Delta m^{2}_{32}=(2.43\pm 0.13)\times 10^{-3}{\rm eV}^{2},\hspace{1cm}\no\\
\sin^{2}(2\theta_{13})&<&0.15,
\end{eqnarray}
using these data, we chose
\begin{eqnarray}\label{angls}
\theta_{12}=\theta\simeq 34.5^{\circ},\hspace{1cm}\theta_{23}=45^{\circ},\hspace{1cm}\theta_{13}=0.
\end{eqnarray}
For the mass squared differences $\Delta m^{2}$, which are given in Eq.~(\ref{data}), energy dependent terms in the probability expression (\ref{prob}), even for high-energy neutrinos ($E_{\nu}\geq1\,\rm GeV$) traveling on cosmologically scales $L\sim\rm Mpc$, are very fast oscillating functions and can be averaged to
\begin{eqnarray}\label{sin}
\sin^{2}\left[3.91\times10^{14}\left(\frac{\Delta m^{2}}{10^{-5}\rm eV^{2}}\right)
\left(\frac{\rm 1\,GeV}{E}\right)\left(\frac{L}{\rm 1\,Mpc}\right)\right]\rightarrow \frac{1}{2}.
\end{eqnarray}
\begin{figure}
\includegraphics[width=4.9in]{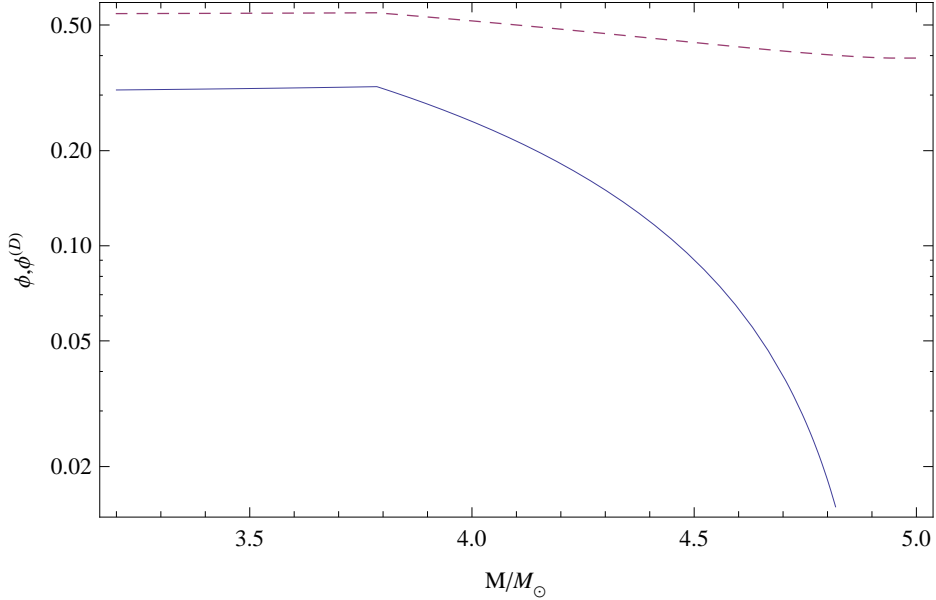}\\
\caption{The quantities $\phi$ and $\phi^{(\rm D)}$ are plotted as a function of collapsing star mass $M$, respectively in solid and dashed lines.}
\label{fig:phi}
\end{figure}
\par
On the basis of results of the Sec.~\ref{sec:nu}, it is clear that primary flux of 10\,MeV neutrinos contains only electron-neutrinos, and their numbers given by Eq.~(\ref{nea}); and primary flux of GeV neutrinos contains electron and muon neutrinos, and their total numbers are given by
\begin{eqnarray}
\mathcal{N}_{\rm e}&=&N_{\bar{\nu}_{\rm e}}^{(b)}+N_{\bar{\nu}_{\rm e}}^{(c)}+N_{\nu_{\rm e}}^{(d)},
\label{nebcd}\\
\mathcal{N}_{\mu}&=&N_{\bar{\nu}_{\mu}}^{(c)}+N_{\nu_{\mu}}^{(d)},\label{nmcd}
\end{eqnarray}
where these numbers can be read from Eqs.~(\ref{neb})-(\ref{nmd}). However oscillation change these numbers, namely produce tau neutrinos which are absent in the primary flux. Using Eqs.~(\ref{prob})-(\ref{sin}) after considering neutrino oscillations, the numbers (\ref{nebcd}) and (\ref{nmcd}) become
\begin{eqnarray}
\mathcal{N}_{\rm e}^{(\rm D)}&=&(1-\frac{1}{2}\sin^{2}(2\theta))\mathcal{N}_{\rm e}+\frac{1}{4}\sin^{2}(2\theta)\mathcal{N}_{\mu},
\label{neo}\\
\mathcal{N}_{\mu}^{(\rm D)}&=&\mathcal{N}_{\tau}^{(\rm D)}\no\\
&=&\frac{1}{2}(1-\frac{1}{4}\sin^{2}(2\theta))\mathcal{N}_{\mu}+\frac{1}{4}\sin^{2}(2\theta)\mathcal{N}_{\rm e}.
\label{nmto}
\end{eqnarray}

Because water Cherenkov detectors, such as Kamiokande, can distinguish electrons from muons but cannot distinguish particles from anti-particles
\cite{Strumia:2006db},
it is useful to investigate, the ratio of the total produced muon neutrinos $\mathcal{N}_{\mu}$ to the total produced electron neutrinos $\mathcal{N}_{\rm e}$ i.e.,
\begin{eqnarray}\label{phi}
\phi\equiv\frac{\mathcal{N}_{\mu}}{\mathcal{N}_{\rm e}}.
\end{eqnarray}
Using Eqs.~(\ref{neo}) and (\ref{nmto}) after considering neutrino oscillations the quantity $\phi$ becomes
\begin{eqnarray}\label{phid}
\phi^{(\rm D)}&\equiv&\frac{\mathcal{N}_{\mu}^{(\rm D)}}{\mathcal{N}_{\rm e}^{(\rm D)}}=\frac{\frac{1}{2}(1-\frac{1}{4}\sin^{2}(2\theta))\phi+\frac{1}{4}\sin^{2}(2\theta)}{1-\frac{1}{2}\sin^{2}(2\theta)+\frac{1}{4}\sin^{2}(2\theta)\phi},
\end{eqnarray}
which shows if $\phi=2$, regardless value of mixing angle $\theta$, flux becomes homogenous i.e., $\phi^{(\rm D)}=1$.
For the flux of 10\,MeV neutrinos and collapsing stars with the mass $M\geq5\,M_{\odot}$, we have $\phi=0$ as a result $\phi^{(\rm D)}=0.39$. Fig.~\ref{fig:phi} shows $\phi$ and $\phi^{(\rm D)}$ for flux of GeV neutrinos, and Fig.~\ref{fig:gev} shows antineutrino events that can be detected on the Earth.  These results can be relevant to observations.

\subsection{\label{sec:det}Detection of neutrinos}

\begin{figure}
\includegraphics[width=4.9in]{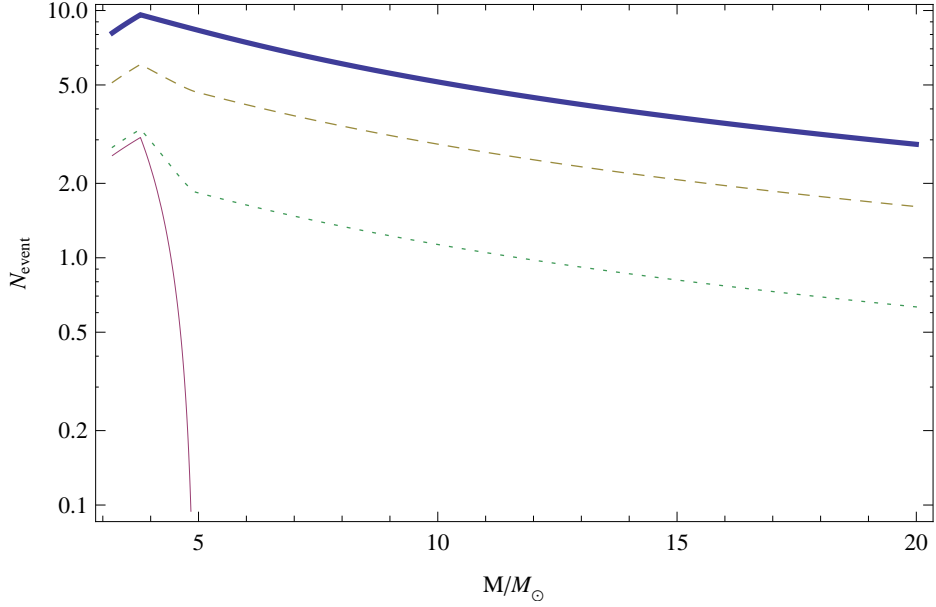}\\
\caption{In a gravitational collapse process, the number of GeV antineutrino events which can be observed by Kamiokande detector is plotted as a function of collapsing star mass $M$. Thick and Thin lines respectively show numbers of electron and muon antineutrino events without considering oscillations. Dashed line shows numbers of electron-antineutrino events by considering oscillations. Dotted line shows number of muon and tau antineutrino events by considering oscillations.}
\label{fig:gev}
\end{figure}
If one is interested only in the charge current processes $\bar{\nu}_{\ell}+p\rightarrow\ell^{+}+n$ and $\nu_{\ell}+n\rightarrow\ell^{-}+p$, so that the neutrino is converted to a charged lepton that can be detected.
\comment{only the reactions $\bar{\nu}_{\ell}+p\rightarrow\ell^{+}+n$ and $\nu_{\ell}+n\rightarrow\ell^{-}+p$ are possible. For the reason that, reactions $\nu_{\ell}+p\rightarrow\ell^{\pm}+n$ and  $\bar{\nu}_{\ell}+n\rightarrow\ell^{\pm}+p$ violate either charge or lepton number conservation.}
From the experimental view point, only the reaction $\bar{\nu}_{\ell}+p\rightarrow\ell^{+}+n$ is interest, because it is not possible to build a target containing enough free neutrons, that would anyway decay \cite{Strumia:2006db}. Enough free protons are obtained using targets made of water $H_{2}O$ such as the Kamiokande detector.
For this reason, analogously to Ref.~\cite{Sekiguchi:2011zd}, we consider only antineutrino events without
taking into account the numbers of 10\,MeV electron-neutrinos, which are produced in the phase $a$, and GeV electron-neutrinos and muon-neutrinos which are produced in the phase $d$.
The number of observed events in a detector depends on the number of emitted neutrinos $N_{\nu}$, cross section of neutrino-nucleon interaction $\sigma$, number of detector target nucleons $N$ and distance from collapsing stars to the Earth $L$. Thus it is determined as \cite{Mohapatra}
\begin{eqnarray}\label{event}
N_{\rm event}=\frac{N_{\nu}}{4\pi L^{2}}\sigma N.
\end{eqnarray}
Here we compute number of GeV antineutrino events, because we are interested in high-energy neutrino detection. The cross section of GeV antineutrinos interaction with a proton approximately is $\sigma\approx10^{-37}{\rm cm}^2$ \cite{Strumia:2006db}. The Kamiokande detector contains $2.2\,\rm KTon$ of water i.e., $N\simeq 1.5\times10^{32}$ hydrogen nucleus or proton \cite{Mohapatra}. For a collapsing star at distance of $L=1\,\rm Mpc$ from the Earth, the number of observed GeV antineutrino events in the Kamiokande detector  $N_{\rm event}\gtrsim 1$, as shown in Fig.~\ref{fig:gev}, depending on collapsing star mass.
\par
\begin{figure}
\includegraphics[width=4.9in]{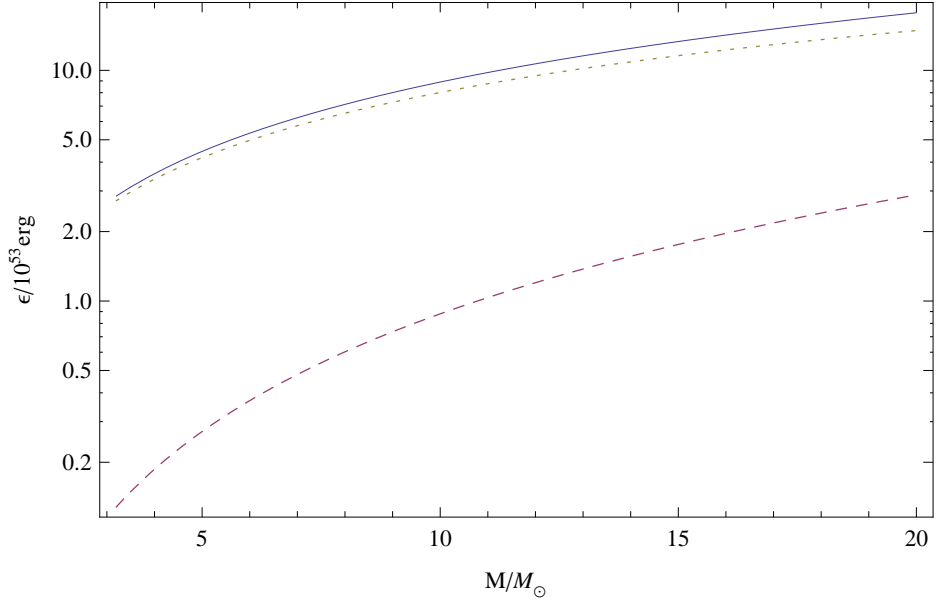}\\
\caption{In unit of $10^{53}\,\rm erg$, the total energy of neutrino flux $\epsilon$ (solid line), the total energy of 10\,MeV electron-neutrino flux $\epsilon_{\rm MeV}$ (dashed line), and the total energy of GeV electron and muon neutrino flux $\epsilon_{\rm GeV}$ (dotted line) are plotted as a function of collapsing star mass $M$.}\label{fig:flux}
\end{figure}

Another important information is the total energy of neutrino flux. Total energy of 10\,MeV electron-neutrinos flux produced in phase $a$ is
\begin{eqnarray}\label{fmev}
\epsilon_{\rm MeV}=N_{\nu_{\rm e}}\times\overline{E}_{\nu_{\rm e}},
\end{eqnarray}
where $N_{\nu_{\rm e}}$ and $\overline{E}_{\nu_{\rm e}}$ are respectively given by Eq.~(\ref{nea}) and shown in Fig.~\ref{fig:ea}. Total energy of GeV electron and muon neutrinos fluxes produced in phases $b,c$ and $d$ is
\begin{eqnarray}\label{fgev}
\epsilon_{\rm GeV}=N_{\bar{\nu}_{\rm e}}^{(b)}\times\overline{E}_{\bar{\nu}_{\rm e}}^{(b)}+(N_{\bar{\nu}_{\rm e}}^{(c)}+N_{\bar{\nu}_{\mu}}^{(c)})\times\overline{E}_{\bar{\nu}_{\rm e},\bar{\nu}_{\mu}}^{(c)}+
(N_{\nu_{\rm e}}^{(d)}+N_{\nu_{\mu}}^{(d)})\times\overline{E}_{\nu_{\rm e},\nu_{\mu}}^{(d)},
\end{eqnarray}
where $N_{\bar{\nu}_{\rm e}}^{(b)},N_{\bar{\nu}_{\rm e}}^{(c)},N_{\bar{\nu}_{\mu}}^{(c)},N_{\nu_{\rm e}}^{(d)},N_{\nu_{\mu}}^{(d)}$ are shown in Figs.~\ref{fig:nb}-\ref{fig:nd} and $\overline{E}_{\bar{\nu}_{\rm e}}^{(b)},\overline{E}_{\bar{\nu}_{\rm e},\bar{\nu}_{\mu}}^{(c)},\overline{E}_{\nu_{\rm e},\nu_{\mu}}^{(d)}$  in Figs.~\ref{fig:eb}-\ref{fig:ed}. Then, the total energy of neutrino flux $\epsilon$ is the sum of $\epsilon_{\rm MeV}$ (\ref{fmev}) and $\epsilon_{\rm GeV}$ (\ref{fgev}). Eqs.~(\ref{fmev}), (\ref{fgev}) and total energy $\epsilon$ of neutrino flux are presented in Fig.~\ref{fig:flux}, which shows the total energy of emitted neutrinos $\epsilon\gtrsim10^{53}\,\rm erg$, depending on collapsing star mass. The total energy of flux presented in Fig.~\ref{fig:flux} can be relevant for observations.

\section{Summary and remarks}

In this article, we present a preliminary study of neutrino productions in gravitational collapse of stellar cores. For this propose we consider the collapsing star as a completely degenerate Fermi gas of neutrons, protons and electrons, because temperature is much smaller than their Fermi energies. To compute neutrino numbers and mean energies one has to solve rate equations, which incorporate both macroscopic hydrodynamic collapsing processes and microscopic particle interactions. However, time scale of macroscopic collapsing processes are much larger than time scale of microscopic particle interactions, as a result collapse can be divided into infinitesimal steps so that at each step the system is in equilibrium state which is given by Eq.~(\ref{cheqil}). By using the chemical equilibrium Eq.~(\ref{cheqil}) and conservation of total electric charge number, total baryon number and total lepton flavor number, number of each species of particles obtained as a function of collapsing star density $\rho$. We investigate the collapsing system from white dwarf density $\rho_{\rm wd}=8.73\times10^{-8}\rho_{\rm nuc}$ to the maximal density at the Schwarzschild radius of the star i.e., $R_{\rm s}=2GM$, at which black hole is formed for stars with the mass $M\geq3.2\,M_{\odot}$. Maximal density $\rho_{\rm s}\equiv M/(\frac{4}{3}\pi R_{\rm s}^{3})$ is plotted as a function of the mass $M$ of collapsing stellar cores in Fig.~\ref{fig:ds} to compare with the density thresholds $\rho_\mu=2.7\,\rho_{\rm nuc}$ and $\rho_\pi=4.6\,\rho_{\rm nuc}$ of muon and pion productions. Based on this figure, there are three cases of collapsing stars. Case 1: stars with the mass $3.2\,M_{\odot}\leq M\leq3.8\,M_{\odot}$ can reach to the density thresholds $\rho_{\mu}$ and $\rho_{\pi}$ of both muon and pion productions. Case 2: stars with the mass $3.9\,M_{\odot}\leq M\leq4.9\,M_{\odot}$ can only reach to the density threshold $\rho_{\mu}$ of muon production. Case 3: stars with the mass $M\geq5\,M_{\odot}$ cannot reach to the density thresholds $\rho_{\mu}$ and $\rho_{\pi}$ of both muon and pion productions. According to these density thresholds there are four phases for neutrino production. The phase $a$ begins from white dwarf density $\rho_{\rm wd}$ to the transition density $\rho_{\rm T}=2.8\times10^{-3} \,\rho_{\rm nuc}$ (\ref{dt}). In this phase, via the inverse $\beta$-decay reaction (\ref{invbta}) electrons and protons having been converted to the neutrons and electron-neutrinos. The Number of these neutrinos is given by Eq.~(\ref{nea}). The phase $b$ begins from the transition density $\rho_{\rm T}$ to the muon production density threshold $\rho_{\mu}$ for cases 1 and 2, and for case 3 to the maximal density $\rho_{\rm s}$. In this phase, via $\beta$-decay reaction (\ref{btadcy}) neutrons are converted into protons, electrons and electron-antineutrinos. The number of these antineutrinos is shown in Fig.~(\ref{fig:nb}). The phase $c$ begins from the muon production density threshold $\rho_{\mu}$ to the pion production density threshold $\rho_{\pi}$ for case 1 and for case 2 to the maximal density $\rho_{\rm s}$. In this phase, in addition to the $\beta$-processes (\ref{invbta}) and (\ref{btadcy}), through reactions (\ref{muon}) and (\ref{mudcy}), electron-antineutrinos and muon-antineutrinos are produced. The numbers of these antineutrinos are shown in Fig.~(\ref{fig:nc}). The phase $d$, which can only be reached in case 1, begins from the pion production density threshold $\rho_{\pi}$ to the maximal density $\rho_{\rm s}$. In this phase, Eqs.~(\ref{nez}) and (\ref{nmz}) show that by increasing the density $\rho$, electron number $N_{\rm e}$ and muon number $N_{\mu}$ decrease. Based on lepton flavor number conservation, electron-neutrinos and muon-neutrinos are produced and their numbers are shown in Fig.~\ref{fig:nd}. The ratio $A/N_{\rm p}$ is an important quantity to show charge composition of compact star as a function of the density $\rho$. As muons and pions produced, this quantity decreases more rapidly as a function of the density $\rho$, as shown in Figs.~\ref{fig:ap}-\ref{fig:apz}. Suppose that 10\% of variation in gravitational energy (\ref{gravty})-(\ref{vargrv}) is converted to neutrino energy, we use the resulted number of neutrino productions (see Sec.~\ref{sec:nu}) and Eq.~(\ref{difme}) to obtain the mean energy of neutrinos. Fig.~\ref{fig:ea} shows that the mean energy of produced electron-neutrinos in the phase $a$ is about 10\,MeV, which is in agreement with neutrino energy observed in supernova explosions. Figs.~\ref{fig:eb}-\ref{fig:ed} respectively show that the mean energy of electron-antineutrinos produced in the phase $b$, electron-antineutrinos and muon-antineutrinos produced in the phase $c$ and electron-neutrinos and muon-neutrinos produced in the phase $d$, are about GeV. The mean energy of total neutrinos and antineutrinos produced in the phases $a,b,c$ and $d$ as shown in Fig.~\ref{fig:et} is about 90\,MeV. The ratio of the total muon neutrinos $\mathcal{N}_{\mu}$ to the total electron neutrinos $\mathcal{N}_{\rm e}$ i.e., $\phi=\mathcal{N}_{\mu}/\mathcal{N}_{\rm e}$ have been computed. For the flux of 10\,MeV electron-neutrinos, which are produced in the phase $a$, and collapsing stars with the mass $M\geq5\,M_{\odot}$ this quantity is zero ($\phi=0$) and by considering the effects of neutrino oscillation becomes $\phi^{(\rm D)}=0.39$ at detectors on the Earth. And for the flux of GeV neutrinos, which are produced in phases $b,c$ and $d$, the quantity $\phi$ ($\phi^{(\rm D)}$) is computed at the source (at detectors on the Earth), the Fig.~\ref{fig:phi} shows that $\phi<1$ ($\phi^{(\rm D)}<1$) at the source (at detectors on the Earth). This result, is completely different from high-energy neutrino production in compact stars without considering star as a completely degenerate Fermi gas which predict the ratio $\phi=2$, see processes (\ref{pd}), at the source and $\phi^{(\rm D)}=1$ after oscillation at the Earth detectors. The number of antineutrino events in an ordinary detector such as Kamiokande is computed for a collapsing star at distance of 1\,Mpc from the Earth. The number of GeV antineutrino events $N_{\rm event}\gtrsim 1$, depending on the mass of collapsing stars, as shown in Fig.~\ref{fig:gev} with and without neutrino oscillation effects. Total energy of neutrino flux $\epsilon\gtrsim10^{53}\,{\rm erg}$, as shown in Fig.~\ref{fig:flux}.
\par
We have to point out that in this approach it is not possible to determine production of neutrinos and antineutrinos via the neutral current interaction, for example $NN$-bremsstrahlung reaction $NN\rightarrow NN+\nu\bar\nu$ \cite{Friman:1978zq}. In addition, an other important assumption in this approach is that as soon as produced, all neutrinos escape away from collapsing stars. By using this assumption, we are able to approximately solve algebraic equations of equilibrium conditions and conservations of particle numbers. In fact, some neutrinos are trapped and participate interactions inside collapsing star before they escape away from collapsing stars, for instance, reactions (\ref{muon}) and (\ref{pimu}). This assumption is on the basis of the chemical potential of trapped neutrinos being smaller than the chemical potential of electrons and muons. If the effects of neutrinos trapped are considered, the number and energy of neutrino emission should be smaller than estimates given in this article. Our results should be considered as  approximate estimations and further numerical calculations are necessary. Nevertheless, the approximate results presented in this article give some physical insight into the neutrino emission in gravitational collapses and show number- and energy-fluxes of neutrinos that could be relevant to observations.




\begin{thebibliography}{99}

\bibitem{Burrows:1987zz}
  A.~Burrows and J.~M.~Lattimer,
  Astrophys.\ J.\  {\bf 318}, L63 (1987).

\bibitem{Arnett:1990au}
  W.~D.~Arnett, J.~N.~Bahcall, R.~P.~Kirshner and S.~E.~Woosley,
  Ann.\ Rev.\ Astron.\ Astrophys.\  {\bf 27}, 629 (1989).

\bibitem{Bethe:1990mw}
  H.~A.~Bethe,
  Rev.\ Mod.\ Phys.\  {\bf 62}, 801 (1990).

\bibitem{Burrows:1990ts}
  A.~Burrows,
  Ann.\ Rev.\ Nucl.\ Part.\ Sci.\  {\bf 40}, 181 (1990).

\bibitem{Hirata:1987hu}
  K.~Hirata {\it et al.}  [KAMIOKANDE-II Collaboration],
  Phys.\ Rev.\ Lett.\  {\bf 58}, 1490 (1987).

\bibitem{Bionta:1987qt}
  R.~M.~Bionta {\it et al.},
  Phys.\ Rev.\ Lett.\  {\bf 58}, 1494 (1987).

\bibitem{Arnett:1976dh}
  W.~D.~Arnett and R.~L.~Bowers,
  Astrophys.\ J.\ Suppl.\  {\bf 33}, 415 (1977).

\bibitem{Cooperstein:1988zz}
  J.~Cooperstein,
  Phys.\ Rev.\  C {\bf 37}, 786 (1988).

\bibitem{Rhoades:1974fn}
  C.~E.~.~Rhoades and R.~Ruffini,
  Phys.\ Rev.\ Lett.\  {\bf 32}, 324 (1974).

\bibitem{Sekiguchi:2011zd}
  Y.~Sekiguchi, K.~Kiuchi, K.~Kyutoku and M.~Shibata,
  Phys.\ Rev.\ Lett.\  {\bf 107}, 051102 (2011)
  [arXiv:1105.2125 [gr-qc]].

\bibitem{Razzaque:2004yv}
  S.~Razzaque, P.~Meszaros and E.~Waxman,
  Phys.\ Rev.\ Lett.\  {\bf 93}, 181101 (2004)
  [Erratum-ibid.\  {\bf 94}, 109903 (2005)]
  [arXiv:astro-ph/0407064].

\bibitem{Razzaque:2005bh}
  S.~Razzaque, P.~Meszaros and E.~Waxman,
  Mod.\ Phys.\ Lett.\  A {\bf 20}, 2351 (2005)
  [arXiv:astro-ph/0509729].

\bibitem{Costantini:2004ap}
  M.~L.~Costantini and F.~Vissani,
  Astropart.\ Phys.\  {\bf 23}, 477 (2005)
  [arXiv:astro-ph/0411761].

\bibitem{Zhang:2002xv}
  B.~Zhang, Z.~G.~Dai, P.~Meszaros, E.~Waxman and A.~K.~Harding,
  Astrophys.\ J.\  {\bf 595}, 346 (2003)
  [arXiv:astro-ph/0210382].

\bibitem{Eichler:1978zp}
  D.~Eichler,
  Nature {\bf 275}, 725 (1978).

\bibitem{Helfand:1979iv}
  D.~J.~Helfand,
  Nature {\bf 278}, 720 (1979).

\bibitem{Dermer:2006xt}
  C.~D.~Dermer,
  J.\ Phys.\ Conf.\ Ser.\  {\bf 60}, 8 (2007)
  [arXiv:astro-ph/0611191].

\bibitem{Stanev:2005kk}
  T.~Stanev,
  J.\ Phys.\ Conf.\ Ser.\  {\bf 39}, 386 (2006)
  [arXiv:astro-ph/0511641].

\bibitem{Becker:2007sv}
  J.~K.~Becker,
  Phys.\ Rept.\  {\bf 458}, 173 (2008)
  [arXiv:0710.1557 [astro-ph]].

\bibitem{Blasi:2000xm}
  P.~Blasi, R.~I.~Epstein and A.~V.~Olinto,
  Astrophys.\ J.\  {\bf 533}, L123 (2000)
  [arXiv:astro-ph/9912240].

\bibitem{Arons:2002yj}
  J.~Arons,
  Astrophys.\ J.\  {\bf 589}, 871 (2003)
  [arXiv:astro-ph/0208444].

\bibitem{Mucke:1999yb}
  A.~Mucke, R.~Engel, J.~P.~Rachen, R.~J.~Protheroe and T.~Stanev,
  Comput.\ Phys.\ Commun.\  {\bf 124}, 290 (2000)
  [arXiv:astro-ph/9903478].

\bibitem{Machner:1999ky}
  H.~Machner and J.~Haidenbauer,
  J.\ Phys.\ G {\bf 25}, R231 (1999).

\bibitem{Halzen:2006mq}
  F.~Halzen,
  Eur.\ Phys.\ J.\  C {\bf 46}, 669 (2006)
  [arXiv:astro-ph/0602132].

\bibitem{mrx2012}
R.~ Mohammadi, R.~ Ruffini and S.-S.~Xue,
arXiv:astr-ph/1206.0431;
W.-B.~Han, R.~Ruffini S.-S.~Xue, arXiv:1110.0700.

\bibitem{Shapiro}
S.~L.~Shapiro and S.~A.~Teukolsky,
(WILEY-VCH Verlag GmbH \& Co. KGaA, Weinheim, 2004).

\bibitem{Weinberg}
S.~Weinberg,
(John Wiley \& Sons, Inc., New York, 1972).

\bibitem{Pagliaroli:2008ur}
  G.~Pagliaroli, F.~Vissani, M.~L.~Costantini and A.~Ianni,
  Astropart.\ Phys.\  {\bf 31}, 163 (2009)
  [arXiv:0810.0466 [astro-ph]].

\bibitem{Ohanian}
H.~C.~Ohanian and R.~Ruffini,
(W.~W.~Norton \& Company, New York, 1994).

\bibitem{Strumia:2006db}
  A.~Strumia and F.~Vissani,
  [arXiv:hep-ph/0606054].

\bibitem{Mohapatra}
R.~N.~Mohapatra and P.~B.~Pal,
(World Scientific Publishing Co. Pte. Ltd., Singapore, 2004).

\bibitem{Nakamura:2010zzi}
  K.~Nakamura {\it et al.} [Particle Data Group],
  J.\ Phys.\ G {\bf 37}, 075021 (2010).

\bibitem{Friman:1978zq}
  B.~L.~Friman and O.~V.~Maxwell,
  Astrophys.\ J.\  {\bf 232}, 541 (1979).

\end{thebibliography}
\end{document}